\documentclass[
	aps,
	prl,
	twocolumn,
	a4paper,
	amsmath,
	amssymb,
	tightenlines,
	notitlepage,
	floatfix,
	nofootinbib
	]{revtex4-2}\pdfoutput=1\pdfsuppresswarningpagegroup=1
\usepackage[utf8]{inputenc}
\usepackage[english]{babel}
\usepackage{bbm}
\usepackage{graphicx}
\usepackage{hyperref}
\usepackage{wasysym}
\usepackage{amsthm}
\usepackage{dcolumn}
	\newcolumntype{.}{D{.}{.}{13}}
	\newcolumntype{d}[1]{D{.}{.}{#1}}
\usepackage[usenames, dvipsnames]{xcolor}
\usepackage[all,cmtip]{xy}
\usepackage{tensor}

\usepackage{ulem}

\graphicspath{{pics/}}

\allowdisplaybreaks[0]
\input{commands.input}

\newcommand{\prlsection}[1]{\paragraph*{#1.\textendash{}}}

\begin{document}

\title{
Intermediate mass-ratio black hole binaries:
Applicability of small mass-ratio perturbation theory
}

\newcommand{\AEI}{\affiliation{Max Planck Institute for Gravitational Physics (Albert Einstein Institute),
    Am M{\"u}hlenberg 1, D-14476 Potsdam, Germany}}

\author{Maarten \surname{van de Meent}}\email{mmeent@aei.mpg.de}\AEI
\author{Harald P. \surname{Pfeiffer}}\AEI

\date{\today}
\begin{abstract}
The inspiral phasing of binary black holes at intermediate
  mass-ratios ($m_2/m_1\sim 10^{-3}$) is important for gravitational wave observations, but
  not accessible to standard modeling techniques: The accuracy of
  the small mass-ratio (SMR) expansion is unknown at intermediate
  mass-ratios, whereas numerical relativity simulations cannot reach this
  regime.   This article assesses the accuracy of the SMR expansion by extracting the first three terms of the SMR expansion from numerical relativity
data  for non-spinning, quasi-circular
binaries.  We recover the leading term predicted by SMR 
theory and obtain a robust prediction of the next-to-leading term.
The influence of higher order terms is bounded to be small, indicating that
the SMR series truncated at next-to-leading order is quite accurate at intermediate mass-ratios and even at nearly comparable
mass binaries.  We estimate the range of applicability for SMR
and post-Newtonian series for non-spinning, quasi-circular inspirals.
\end{abstract}

\maketitle \setlength{\parindent}{0pt} \setlength{\parskip}{6pt}

Inspiraling and merging black hole (BH) binaries are the most numerous
source of gravitational waves(GW) observed by the LIGO and Virgo
detectors~\cite{TheLIGOScientific:2014jea,TheVirgo:2014hva}
and are one of the key science targets for third generation ground-based GW detectors~\cite{3GScienceCase}, as well as the space-based LISA
observatory~\cite{Audley:2017drz}.  The mass-ratio $q\equiv m_2/m_1\le 1$ is one of the key
parameters in the dynamics of these systems.  The LIGO and Virgo
observations~\cite{LIGOScientific:2018mvr,LIGOScientific:2018jsj,Venumadhav:2019lyq} mostly report $q$ close to
unity,
with GW190412~\cite{LIGOScientific:2020stg} and GW190814~\cite{Abbott:2020khf} the first systems with
clearly unequal masses ($q\sim 0.28$ and $q\sim 0.11$).

In the future, observations of binaries with lower $q$ are
expected:
Continued observations with the current detectors~\cite{Aasi:2013wya} may reveal binaries with smaller $q$.
Third generation ground based detectors with improved low
frequency sensitivity will be able to detect the capture of stellar
mass BHs by intermediate mass BHs with mass-ratios
down to $q\sim 10^{-3}$~\cite{Jani:2019ffg}.
LISA will observe the mergers of
massive BHs of millions of solar masses. While the
majority of these are expected to have $q\gtrsim 0.1$, there could a
significant tail of events down to $q\sim
0.01$~\cite{Salcido:2016oor,Volonteri:2020wkx}.
LISA will also be sensitive to mergers of intermediate mass BHs with massive BHs
($q\sim 10^{-3}$) and extreme mass-ratio inspirals
($q\sim10^{-5}$) as sensitive probes of black hole
physics~\cite{Audley:2017drz}.

The modeling of inspiraling binaries at all mass-ratios is therefore of paramount importance for
detection and analysis of GW sources.  The three primary modeling approaches  are post-Newtonian slow-velocity perturbation
theory~\cite{Blanchet:2013haa}, numerical relativity (NR), i.e.\ direct numerical integration of the full non-linear Einstein
equations~\cite{Baumgarte:2010ndz}, and small mass-ratio (SMR) perturbation
theory~\cite{Poisson:2011nh}. Effective one body methods~\cite{Buonanno:1998gg}
provide a means to combine and resum information from all three approaches and
also from newer developments like post-Minkowski
expansions~\cite{Damour:2016gwp}.

This article examines whether the SMR and NR approaches combined can  accurately model binaries with any mass-ratio or whether there is a `gap' at intermediate mass-ratios where neither SMR nor NR is sufficiently accurate.  The SMR
approximation expands the dynamics of a coalescing binary in powers of
$q$ or the symmetric mass-ratio $\nu\equiv m_1m_2/(m_1+m_2)^2= q 
+ \bigO(q^2)$.
At leading order, the secondary object follows a geodesic in the background space-time
generated by the primary.
The impact of the secondary's mass on the
dynamics can be included as an effective force term, the gravitational
self-force (GSF).
Calculation of the GSF has progressed rapidly over the past two decades (see \cite{Barack:2018yvs} for a review), but the full next-to-leading order contribution to the orbital phasing has not yet been obtained.  While the main motivation for SMR lies in extreme-mass-ratio
inspirals, there is increasing
evidence~\cite{LeTiec:2011bk,Sperhake:2011ik,LeTiec:2011dp,Nagar:2013sga,Tiec:2013twa,LeTiec:2017ebm,vandeMeent:2016hel} that the
SMR may be applicable even at comparable masses.

Numerical relativity directly solves the full non-linear Einstein
equations~\cite{Baumgarte:2010ndz}.  The vast
majority of simulations performed to date are at comparable masses, with
only very few simulations at $q\lesssim 0.1$ (see,
e.g.~\cite{Boyle:2019kee}, but
note~\cite{Husa:2015iqa,Lousto:2020tnb} for simulations at $q=1/18$ to
$q=1/128$).  The limited coverage in $q$ has two causes.  First, the
number of orbits the binary spends in the strong field region grows
$\propto \nu^{-1}$.  Second, because of
the Courant-limit on the time-step of the numerical simulations, the
number of time-steps per orbit increases $\propto q^{-1}$.  Combined,
these effects cause an increase in computational cost at least
quadratically in mass-ratio.  The need for higher numerical resolution
to resolve the ever smaller secondary (as $q\to 0$), and to preserve
phase-accuracy over the increasingly longer inspiral 
will increase computational cost further.

Given the expectation of binaries at all mass-ratios, the
question arises how to model intermediate mass-ratio binaries at small
separation: post-Newtonian theory is not accurate close to merger owing to the high velocities; numerical relativity simulations are
limited to large mass-ratios, $q\gtrsim 0.1$; and the SMR
approximation is presently only available at leading order in $q$,
and thus may be inaccurate at intermediate mass-ratios.
This letter investigates the 
existence of a mass-ratio
gap where none of the modeling approaches is applicable.  We analyse
NR simulations at mass-ratios $0.1\le q\le 1$ computed with the  
\texttt{SpEC}-code~\cite{Boyle:2019kee,Mroue:2013xna} and extract the first three
terms in the SMR expansion of the orbital phasing.  Analysing these
terms, we conclude that SMR results at next-to-leading order
can likely bridge the mass-ratio gap up to
mass-ratios $q$ large enough to be covered by numerical relativity.

\prlsection{Methodology}

We use geometric units such that $c=G=1$ and examine the orbital phase extracted from the gravitational
radiation at future null infinity,
\begin{equation}\label{eq:phi}
\phi \equiv \frac{1}{2} \arg h_{22}.
\end{equation}
Here $h_{22}$ is the spin-weight $s=-2$ spherical harmonic
$(\ell,m)=(2,2)$ mode of the complex GW strain.
The current work focuses on non-precessing
binaries where Eq.~(\ref{eq:phi}) is sufficient. 

Introducing the orbital frequency,
\begin{equation}\label{eq:Omega}
\Omega \equiv  \d{\phi}{t},
\end{equation}
we consider the orbital phase as a function of the orbital frequency,
$\phi(\Omega)$.
In the SMR approximation, $\phi$ can be calculated by a two
timescale expansion \cite{Hinderer:2008dm} leading to a power series
in the mass-ratio, known as the post-adiabatic (PA) expansion,
\begin{equation}\label{eq:PAexp}
\phi(\Omega) = \sum_{n=0}^{\infty} \nu^{n-1} \phi_{n\rm PA}(M\Omega).
\end{equation}
Here, $\phi_{n\rm PA}$ are functions of $M\Omega$, where
$M\equiv m_1+m_2$ is the total mass of the binary.  Alternatively, one
can consider
$\phi_{n\rm PA}$ as functions of $m_1\Omega$ and/or expand in $q$ as the small parameter (cf.~Fig.~\ref{fig:1PA}
below).

The leading order term~\cite{Hinderer:2008dm,vandeMeent:2018rms} $\phi_{0\rm PA}$ (called ``adiabatic'' or ``0-post-adiabatic'') 
is independent of the
choice of expansion parameter or mass-normalization.  It can be 
computed by
energy balance,
\begin{equation}\label{eq:energybalance}
 \d{\phi_{0\rm PA}}{\Omega}= \nu\,\Omega 
 \d{E}{\Omega}\left(\d{E}{t}\right)^{-1},
\end{equation}
where $E(\Omega)$ is the specific energy of the circular geodesic with
orbital frequency $\Omega$, and ${\rm d}E/{\rm d}t$ its energy loss to
GWs.  We compute  ${\rm d}E/{\rm d}t$
with
the \textsc{Black Hole Perturbation Toolkit}~\cite{BHPToolkit}, utilizing the arbitrary precision Teukolsky code developed in
\cite{Fujita:2004rb,Fujita:2009us,Throwethesis,vandeMeent:2014raa,vandeMeent:2015lxa,vandeMeent:2016pee,vandeMeent:2017bcc}, and denote the result as $\phi_{0\rm PA}^{\rm SMR}$ below.

The 1PA term in the expansion requires knowledge of the full
first-order GSF for nearly circular orbits, and the dissipative part
of second-order GSF for quasi-circular inspirals \cite{Hinderer:2008dm,vandeMeent:2018rms}. Calculation of the
first order GSF for non-spinning binaries is now routine
\cite{Barack:2007tm,Barack:2010tm,Akcay:2013wfa,Osburn:2014hoa,Merlin:2014qda}.  The
calculation of second order GSF for quasi-circular orbits, however,
remains an open challenge in GSF theory, although steady progress has
been
made~\cite{Pound:2012nt,Pound:2012dk,Pound:2014xva,Pound:2014koa,Pound:2015wva,Miller:2016hjv,Pound:2017psq,Pound:2019lzj}.

We use numerical relativity simulations from the \texttt{SpEC}-code, which utilizes
the quasi-local angular momentum formalism to monitor the black hole
spins~\cite{Cook:2004kt,Caudill:2006hw,Cook:2007wr,Lovelace:2008tw},
iterative eccentricity reduction to achieve orbital eccentricities
$e\lesssim 10^{-4}$~\cite{Pfeiffer:2007yz,Buchman:2009ew}, and solves
the Einstein evolution equations in the generalized harmonic
formulation~\cite{Friedrich1985, Garfinkle:2001ni, Pretorius:2004jg,
  Lindblom:2005qh} with constraint damping and minimally reflective
outer boundary conditions~\cite{Lindblom:2005qh, Rinne:2006vv,
  Rinne:2007ui} (see~\cite{Boyle:2019kee} for more details).  Because
of the use of spectral methods and a dual-frame
approach~\cite{Scheel:2006gg} \texttt{SpEC} achieves very high
accuracies even for long inspiral simulations that cover a
comparatively large range in orbital frequencies.  Gravitational
radiation is extracted using the Regge-Wheeler-Zerilli formalism,
extrapolated to future null
infinity~\cite{Boyle:2009vi,Boyle:2019kee}, and corrected for center
of mass drifts~\cite{Woodford:2019tlo}.

This study utilizes 55 NR simulations
of non-spinning quasi-circular inspirals from the public SXS catalog
\cite{SXSCatalog2019,SXSCatalog} with mass-ratios $q\in[0.1,1]$.  The initial orbital frequency is in the range
$M\Omega\sim 0.015 \ldots 0.02$.  Simulations with smaller $q$ tend to
start at the higher frequencies, to achieve a computationally
manageable overall duration of the simulations.  All simulations
are available at multiple numerical resolutions for convergence tests.

The orbital phase $\phi^{\rm NR}(M\Omega)$ is determined by locally fitting a
low order polynomial in $t$ to $\phi^{\rm NR}(t)$. The width of the fitting
window is variable such that at low frequencies it encompasses several
radial oscillations of any residual eccentricity in the simulations,
while at larger frequencies it is small enough to avoid systematic
bias due to the rapidly changing frequency.  The constant of
integration when integrating Eq.~(\ref{eq:Omega}) is chosen such that
$\phi=0$ at $M\Omega=0.046$. 
At a given value of $M\Omega$, the post-adiabatic coefficients $\phi_{n\rm PA}(M\Omega)$ are
determined by fitting a polynomial in $\nu$ to the
data-points $(\nu_A, \phi^{\rm NR}_A(M\Omega))$, where $A=1,\ldots, 55$ labels
the NR simulations, and $\nu_A$ is the symmetric mass-ratio of each
simulation.  This fit is repeated for many values
of $M\Omega$.  Error estimates are obtained by repeating this
procedure with (i) medium-resolution NR simulations; (ii) using the
Weyl-scalar $\Psi_4$ instead of the  GW strain in
Eq.~(\ref{eq:phi}); (iii) vary the order with which the GW strain is extrapolated to future null infinity; and (iv) changing the number of terms in the fit
of form Eq.~(\ref{eq:PAexp}) between three and four.
The range of these calculations is reported as error bar in our results.
At each frequency,
only those NR simulations are used that have a starting frequency
below; for $M\Omega\lesssim 0.02$, the reduced number of available NR
simulations causes larger error bars.

\prlsection{Results}

\begin{figure}
\includegraphics[width=0.97\columnwidth,trim=0 7 0 0]{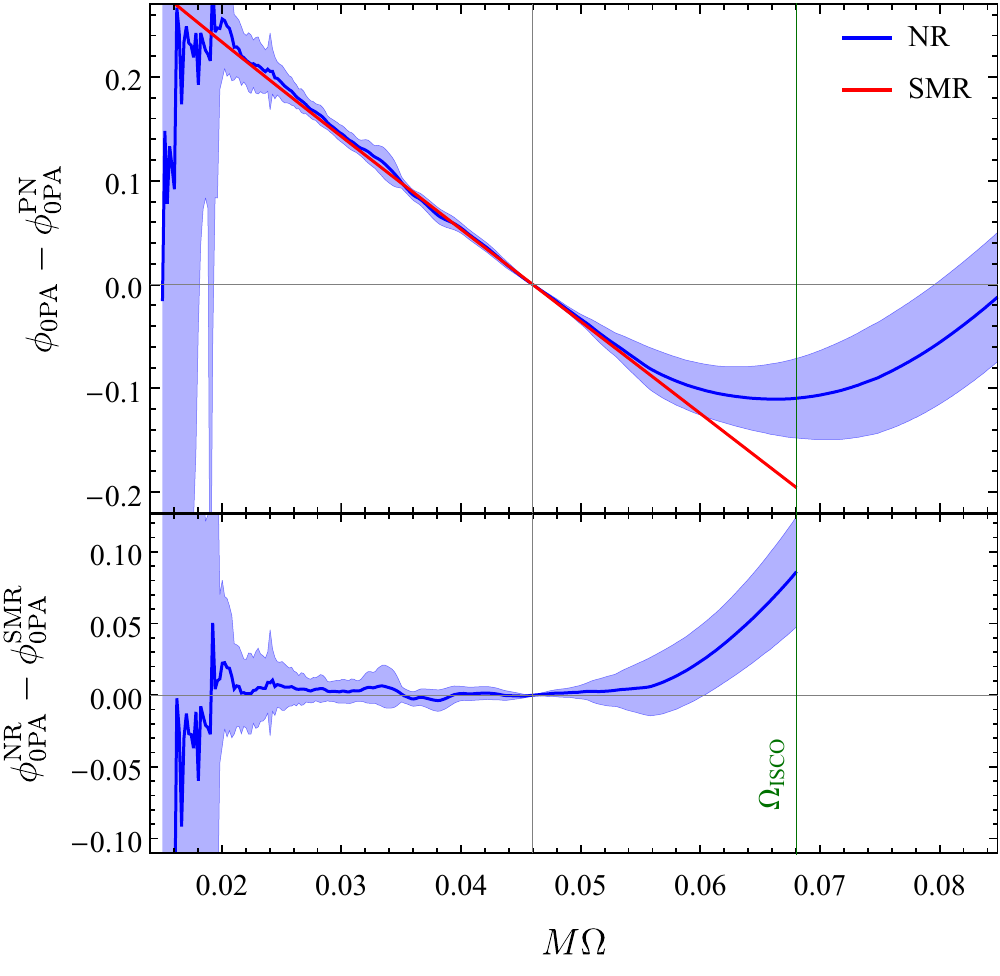}
  \caption{\label{fig:0PA} \textbf{Top:} Leading order in mass-ratio
    contribution to the orbital phasing of the quasi-circular inspiral
    of non-spinning black holes.  Shown are the result derived here from NR simulations ('NR'), as well as the
    small-mass-ratio perturbation theory ('SMR').  For both curves,
    the 3.5PN result~\cite{Blanchet:2013haa} was substracted for
    clarity of plotting.
\textbf{Bottom:} Difference between SMR and the NR result. The shaded areas indicate the estimated uncertainty of the numerical calculation and $\Omega_{\rm ISCO}$ indicates the last stable orbit for $\nu=0$.}
\end{figure}

The leading order term $\phi_{0\rm PA}(M\Omega)$ can be extracted
with good accuracy from the NR simulations, as shown in
Fig.~\ref{fig:0PA}.  To reduce the dynamic range on the y-axis, this
figure shows the difference to the post-Newtonian $\phi_{0\rm PA}^{\rm
  PN}$ result at order $(v/c)^7$, taken from \cite{Blanchet:2013haa}.
The blue curve represents the result of our analysis of NR simulations
(with error bar), whereas the red line is the leading order SMR result
computed by Eq.~(\ref{eq:energybalance}).
The agreement between the two is quite remarkable, and is a first
indication that the PA expansion of the phase in the mass-ratio is
well-behaved for comparable mass-ratios.

At higher frequencies, $M\Omega\gtrsim 0.055$, we find an apparently
systematic deviation between NR and the SMR result.  This deviation
may arise from a breakdown of the PA expansion near the last stable
orbit as the binary transitions from inspiral to plunge.  Studies of
this transition regime~\cite{Buonanno:2000ef,Ori:2000zn} lead to order
$\nu^{-1/5}$ corrections to Eq.~\eqref{eq:PAexp}. Including such a
term in our fit does indeed eliminate the systematic deviation at
$M\Omega\gtrsim 0.055$.  However, the additional term is nearly
degenerate with the 0PA and 1PA terms at low frequencies making it
impossible to get robust numerical results for $\phi_{1\rm PA}$ and
higher. Therefore, we proceed in our analysis without such transition
terms.

\begin{figure}
  \includegraphics[width=0.98\columnwidth,trim=0 7 0 0]{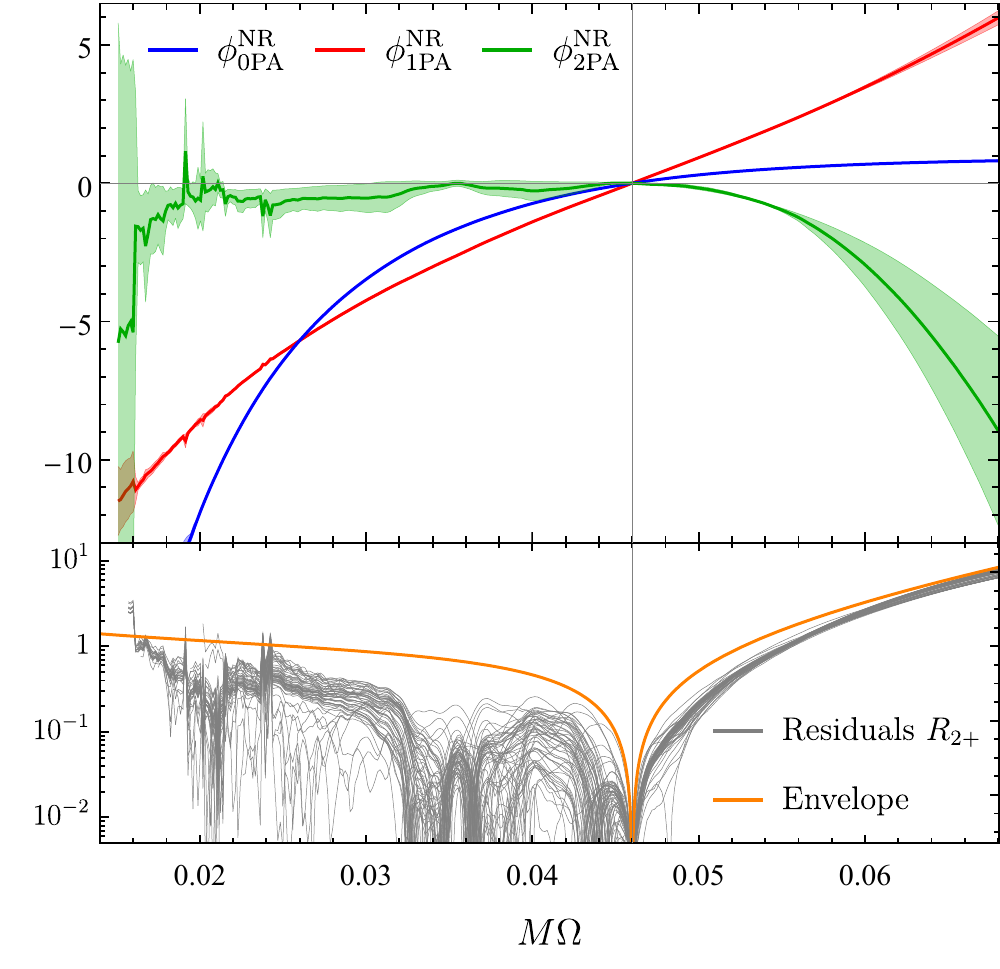}
  \caption{\label{fig:0PA+1PA+2PA} \textbf{Top:} The three leading
    terms in the mass-ratio expansion of the orbital phase, as
    computed here. \textbf{Bottom:} Residuals $R_{2+}$ for all 55 NR
    simulations indicating the combined contributions of 2PA and
    higher, as well as an envelope bounding these residuals in a
    $\nu$--independent manner.  }
\end{figure}

Given how well the numerically extracted 0PA term agrees with its SMR
prediction, we henceforth set it to the SMR value when fitting for the higher
order PA terms.  Figure~\ref{fig:0PA+1PA+2PA} shows the 1PA and 2PA
term obtained from the NR simulations, together with the 0PA term
already discussed in Fig.~\ref{fig:0PA}.
The coefficients $\phi_{n\rm PA}$ are of comparable magnitude in the
frequency range covered by our analysis, suggesting that the PA series
is convergent at equal masses.  Moreover, for frequencies
$M\Omega\lesssim 0.05$, the 2PA coefficient is almost consistent with
zero, i.e.\ the 0PA and 1PA terms already capture essentially all
variation  due to mass-ratio in the numerical data at these frequencies.  In fact,
``goodness-of-fit'' indicators, such as the adjusted $R^2$ value, show
only marginal improvements when adding terms to the fit beyond the 1PA
coefficient.

\begin{figure}
\includegraphics[width=0.92\columnwidth,trim=0 7 0 1]{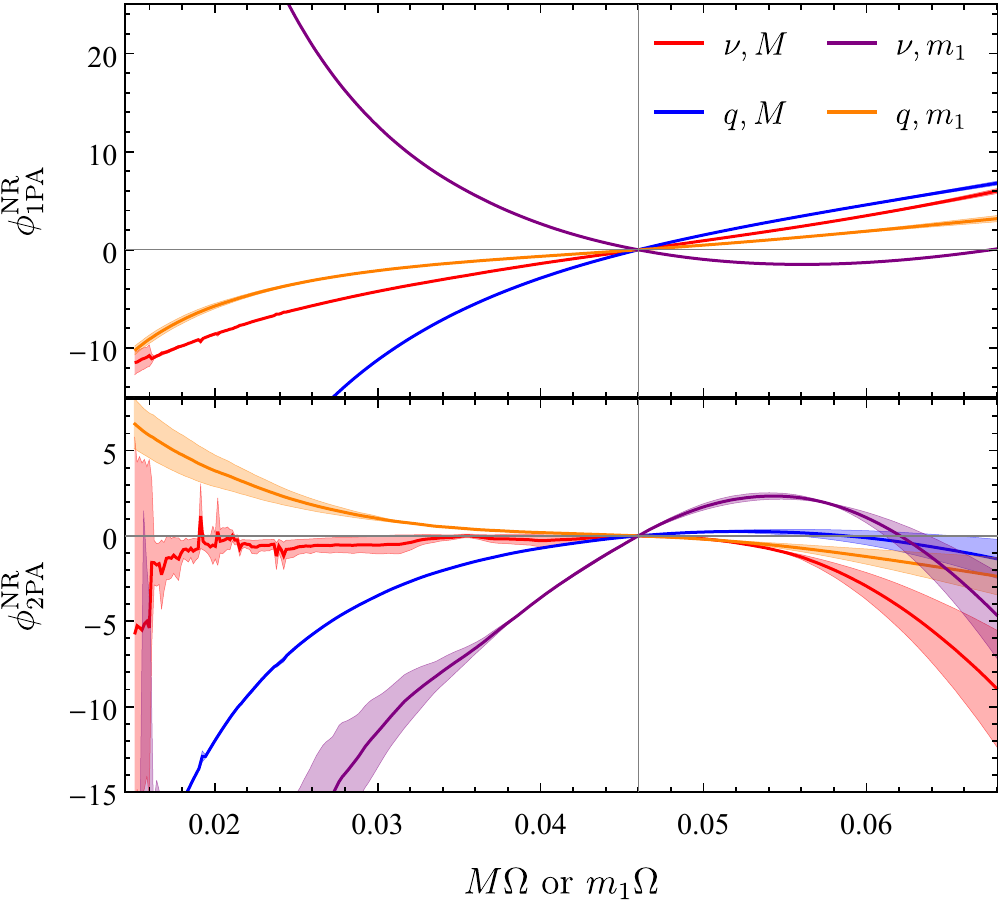}
 \caption{\label{fig:1PA} Impact of the choice of expansion parameters
   on the 1PA and 2PA contributions to the orbital phasing.  The
   different curves differ in whether in Eq.~(\ref{eq:PAexp}) is
   expanded in powers of $\nu$ or in $q$, and whether the
   $\phi$-functions are written in terms of $M\Omega$ or $m_1\Omega$.
   The combination $\nu, M\Omega$ yields an exceptionally small 2PA
   term at low frequencies.
 }
\end{figure}

\begin{figure}
\includegraphics[width=\columnwidth,trim=2 7 2 4]{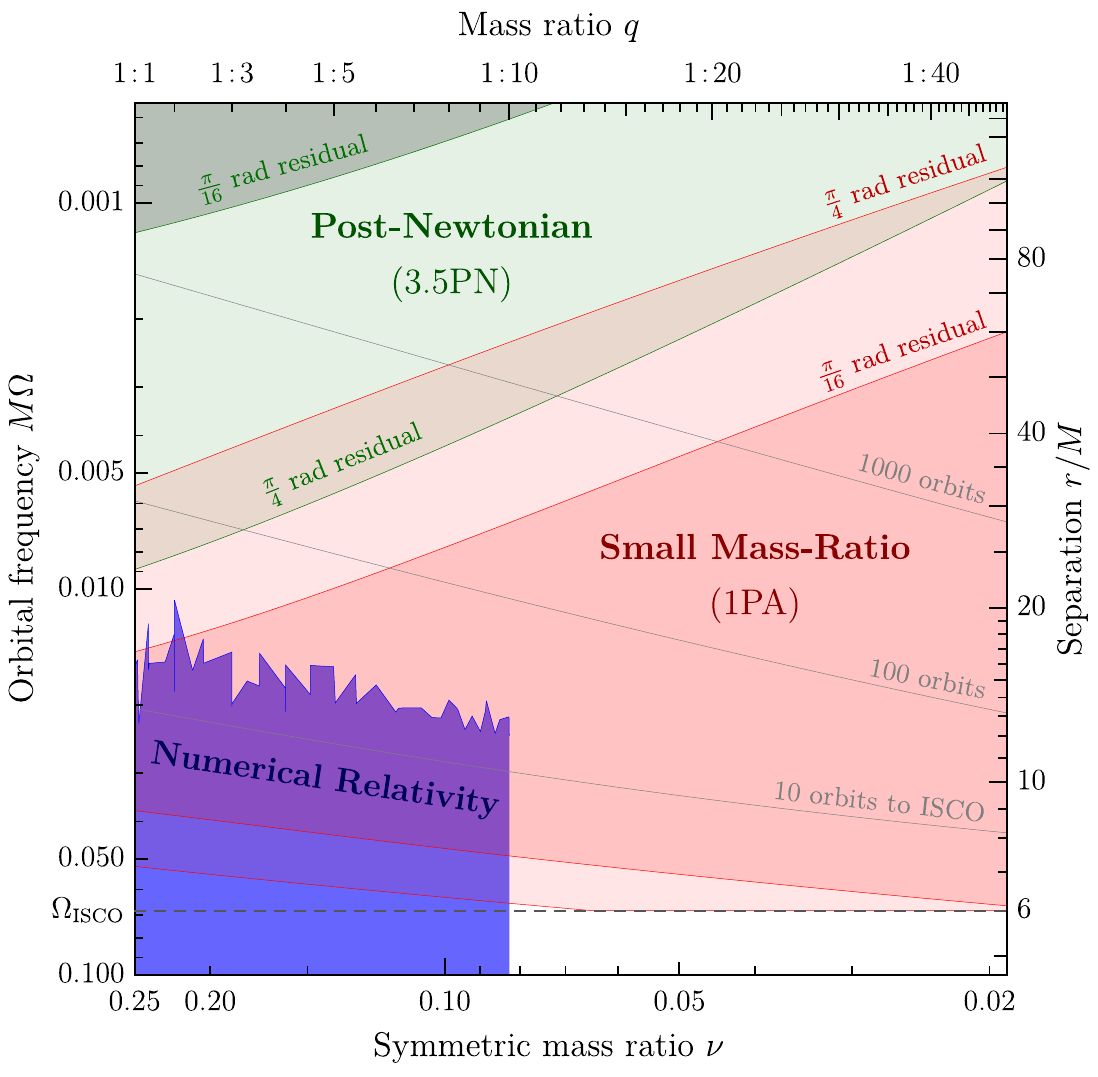}
\caption{\label{fig:phasediagram} Region of applicability of different
  approximation techniques for non-spinning quasi-circular binary
  black hole inspiral.  The shaded regions indicate ranges within
  which the cumulative orbital phase-error is less than $\pi/4$ and $\pi/16$ radians, respectively.  }
\end{figure}

The lower panel of Fig.~\ref{fig:0PA+1PA+2PA} provides a different view
on the importance of terms beyond $\phi_{1\rm PA}$: For each of the 55 NR simulations, this panel plots
\begin{equation}
  R_{2+}\equiv \frac{1}{\nu}\left(\phi^{\rm NR}-\frac{1}{\nu}\phi_{0\rm PA}-\phi_{1\rm PA}^{\rm NR}\right),
  \end{equation}
i.e.\ the contribution of all terms $n\ge 2$ in Eq.~(\ref{eq:PAexp}),
with overall $\nu$ scaling compensated.
All $R_{2+}$ can be
bounded independent of mass-ratio by an envelope function, consisting of the known 3.5PN terms
of $\phi_{2\rm PA}$ and a higher order polynomial in $M\Omega$ fitted by
eye.

So far, we have expanded in symmetric mass-ratio $\nu$, while scaling
orbital frequencies by total mass $M$, cf.\ Eq.~(\ref{eq:PAexp}).  One
can also use the mass-ratio $q=m_2/m_1$ as the small parameter, and/or
scale orbital frequency by the large body's mass $m_1$.  This yields
four variations, all of which agree at the leading 0PA-order.
Figure~\ref{fig:1PA} presents the results for the 1PA and 2PA
contributions. In all four cases, the extracted 1PA and 2PA
coefficients remain of similar magnitude, implying that the expansion
is not dominated by higher order terms. However, the 2PA term is
remarkably small only when expanding using the symmetric mass ratio
$\nu$ and total mass $M$.  The choice $\nu, M$ is indeed preferred as
it is invariant under exchange of the two bodies
$1\leftrightarrow 2$~\cite{LeTiec:2011bk,Tiec:2014lba}.

\prlsection{Discussion} The phasing of inspiraling BH binaries
  is of utmost importance for GW astronomy to find signals, determine
  their parameters and to perform tests of general relativity.
  Binaries at intermediate mass-ratios $q\sim 10^{-3}$ are in a regime
  not accessible to NR, while potentially out of reach for SMR perturbation
  theory. This situation is compounded by the difficulty of
  calculations of the SMR expansion, for which today only the leading
  order (called the zeroth post-adiabatic order) is fully known.
  Here, we extract the first three terms of the SMR-expansion
  from NR simulations at comparable masses, $q\ge 0.1$,
  and use these results to perform  the first comparison between NR and SMR
expanded results for a gauge invariant quantity that includes both
dissipative and conservative effects, namely the accumulated orbital
phase as a function of orbital frequency $\phi(M\Omega)$. We have
successfully extracted the post-adiabatic expansion of this quantity
as a power series in the mass-ratio from non-spinning quasi-circular
NR simulations.

The leading adiabatic (0PA) term agrees with the
result from SMR calculations. In addition we obtain a robust
determination of the 1PA term, serving as a concrete prediction for
the ongoing SMR calculation of this term, which requires the dissipative part
of the second order gravitational self-force.
We also estimate the 2PA term $\phi_{2\rm PA}$ from the NR data. Its amplitude is comparable to $\phi_{0\rm PA}$ and $\phi_{1\rm PA}$ for the frequency-range considered here, indicating that the PA-expansion remains well-behaved. 
In particular, when the PA series is expanded in
powers of the symmetric mass-ratio while keeping the total mass fixed,
the 2PA and higher order terms are consistent with zero within the numerical accuracy for $0.015\lesssim M\Omega\lesssim 0.05$.
For higher frequencies (approaching the last stable circular orbit),
we find indications of a transition-regime to plunge where the series
in integer powers of $\nu$ is no longer applicable.

Our analysis allows us to delineate the regions of applicability of
SMR, NR and PN in a quantitative way, as shown in
Fig.~\ref{fig:phasediagram}: Assuming $\phi_{1\rm PA}$ will become
available through GSF calculations, the envelope to the $R_{2+}$ in
Fig.~\ref{fig:0PA+1PA+2PA} gives a bound on the secular contributions
of higher PA terms.  The red shaded areas in
Fig.~\ref{fig:phasediagram} show the largest $M\Omega$ interval that
can be covered such that the
total accumulated phase-error due to
$\geq 2$PA terms is below a certain value.  The region of
applicability of SMR increases toward smaller mass-ratios, but is
still non-negligible even at comparable masses.  The post-Newtonian
errors are estimated by fits against $\phi^{\rm NR}(M\Omega)$,
cf.\ top panel of Fig.~\ref{fig:0PA}.  The green shaded areas indicate
regions where the cumulative 3.5-PN phase-error for the entire
inspiral up to the given frequency is below a certain value.  Finally,
the blue shaded area indicates the region covered by the NR
simulations used here.  These simulations have phase-accuracy better
than the $\pi/16$ contour line, indicating that the usability of NR is
not limited by accuracy but rather by the length of the simulations.
The three modelling approaches deliver complementary information,
covering different regions of the parameter space.  The region of
validity of each method depends on the desired accuracy, and it also
depends on the use of the waveforms: For GW astronomy, only the
accuracy within the frequency band
of the relevant GW detectors is important, and this will depend on
the total mass of the binary.
Moreover, the needed accuracy will 
depend on the signal-to-noise ratio
at which it is observed.

We note that the adiabatic $\phi_{0\rm PA}$ term is never accurate
enough in the metric of Fig.~\ref{fig:phasediagram}, because
$\phi_{1\rm PA}$ contributes tens of radians in the frequency range
considered, independent of the mass-ratio.  This underlines the
importance of calculating the 1PA term (and therefore the second order
gravitational self-force) for modelling binaries of any mass-ratio.
Furthermore, the application of the 1PA approximation for low
frequencies is limited by a $(M\Omega)^{-1/3}$ divergence of the 2PA term. This
motivates the development of models that incorporate both SMR and PN results,
e.g. using effective-one-body theory~\cite{Antonelli:2019fmq,Damour:2009sm,Akcay:2012ea,Akcay:2015pjz}.

The results in this paper come with two important caveats. First, our
results are limited to non-spinning quasi-circular black hole
binaries.  Adding spin or eccentricity makes the waveform considerably more complex and could make the convergence of
the PA series significantly worse.  Future studies are needed to
explore the full parameter space.
Even for non-spinning quasi-circular case, NR simulations at smaller mass-ratio
are needed to investigate the transition to plunge, as well as longer simulations, to extend our analysis to smaller frequencies.

Second, the current analysis applies only to the inspiral, since the
PA expansion is known to breakdown at the last stable orbit. Our results
motivate the development of 1PA accurate models that also include
plunge, merger, and ringdown, as has previously been done at 0PA order~\cite{Rifat:2019ltp}.

\begin{acknowledgments}
  \prlsection{Acknowledgments}
  The authors acknowledge use of public NR data from the SXS
  Collaboration~\cite{SXSCatalog2019}.
\end{acknowledgments}

\raggedright \bibliography{journalshortnames%
	,meent%
	,commongsf%
	,PAfromNR%
}

\end{document}